\documentclass[aps,prx,showpacs,twocolumn,amsmath,amssymb]{revtex4-1}
\usepackage{graphicx}
\usepackage{dcolumn}
\usepackage{bm}
\usepackage{amsmath}
\usepackage{amsfonts}
\usepackage{amssymb}
\usepackage{multirow}

\usepackage{amsthm}
\usepackage{pinlabel}
\usepackage{natbib}

\usepackage{epstopdf}
\usepackage[abs]{overpic}

\usepackage{cancel}

\usepackage{lipsum}  

\usepackage[normalem]{ulem}

\usepackage[dvipsnames]{xcolor}

\usepackage{xr}
\newcommand{\no}{\nonumber}

\usepackage{braket}

\definecolor{specialgray}{HTML}{505050}
\definecolor{col10K}{HTML}{FFA000}
\definecolor{col300K}{HTML}{924FA4}
\definecolor{colMu}{HTML}{5278BD}
\definecolor{colMuI}{HTML}{924FA4}
\definecolor{newred}{HTML}{D53E4F}
\definecolor{newblue}{HTML}{5278BD}
\definecolor{newcyan}{HTML}{4EBCB3}
\definecolor{newgreen}{HTML}{5CB14E}
\definecolor{newpurple}{HTML}{924FA4}
\definecolor{newyellow}{HTML}{D1C72E}
\definecolor{neworange}{HTML}{D6923C}

\usepackage{lineno}
\usepackage{placeins}

\begin{document}
\title{Cascade of Replica Bands in Flat Band Systems:  Predictions for Twisted Bilayer Graphene}
\author{Fabian Schrodi}\email{fabian.schrodi@physics.uu.se}
\author{Alex Aperis}\email{alex.aperis@physics.uu.se}
\author{Peter M. Oppeneer}\email{peter.oppeneer@physics.uu.se}
\affiliation{Department of Physics and Astronomy, Uppsala University, P.\ O.\ Box 516, SE-75120 Uppsala, Sweden}
	
\vskip 0.4cm
\date{\today}

\begin{abstract}
	\noindent 
	We investigate the effect of electron-phonon interactions (EPI) in systems exhibiting one or more flat electron bands close to the Fermi level and a comparatively large phonon energy scale. After solving the self-consistent full-bandwidth Eliashberg equations, we compute angular resolved photoemission spectroscopy (ARPES) and scanning tunneling spectroscopy/microscopy (STS/STM) spectra. We obtain a sequence of quasiparticle replica bands in both the normal and superconducting states that originate from frequency dependent features of the electron mass renormalization function. We show that these replica  bands can be used to extract the relevant phonon energy scale from experiments. Focusing in particular on twisted bilayer graphene, we predict replica-band formation which, when observed, will shed light on the role of EPI in this archetypal flat-band system. 
\end{abstract}

\maketitle

\textit{Introduction.} 
Effects of electron-phonon interactions in metals and superconductors are most accurately modeled by the Eliashberg formalism\,\cite{Migdal1958,Eliashberg1960}. The connection between this theory and STS/STM spectra
has been well understood for many decades\,\cite{Rowell1963,Schrieffer1963}. In more recent years, advances in ARPES have lead to the possibility of an even richer comparison between theory and experiment\,\cite{Vekhter2003,Shi2004,Cuk2005,Chien2009}. These techniques have been successfully applied to gain better understanding of many materials, such as the high-temperature superconducting cuprates\,\cite{Cuk2004,Devereaux2004,Gweon2006,Bok2016} and monolayer FeSe on a SrTiO$_3$ (STO) substrate\,\cite{Rademaker2016,Aperis2018,Schrodi2018}, to name only few examples.

In this work we focus on systems with one or more flat electron energy bands close to the Fermi level, where the term `flatness' is to be understood in comparison to the phonon energy scale $\Omega$. Calculating ARPES and STS/STM spectra using full-bandwidth Eliashberg theory, we reveal a sequence of quasiparticle replica bands outside the electron bandwidth $W$ of the flat bands, occurring both at positive and negative frequencies. A closer analysis reveals that these spectral features are a direct manifestation of the electron mass renormalization function, and they are located at integer multiples of $\Omega$ along the frequency axis. Therefore they can serve as a means to extract the phonon frequency directly from the measured spectra.  Notably, in contrast to the replica bands observed for FeSe/STO \cite{Lee2014} which are mediated by the forward-scattering form of the EPI \cite{Rademaker2016,Aperis2018},  our here predicted cascade of replicas is driven solely by the flatness of the electron band and therefore has a distinct origin which is more akin to the  shake-off spectral features of localized systems \cite{mah00}.

We apply our theory to twisted bilayer graphene (TBG) at the magic angle $\sim1.1^{\circ}$  \cite{Cao2018_1,Cao2018_2,Yankowitz2019}, for which we perform calculations of the quasiparticle spectrum using material specific input. This system exhibits two flat bands close to the Fermi energy, as reported by both theory\,\cite{dosSantos2007,Bistritzer2011,Haddadi2020} and experiment\,\cite{Li2010,Kerelsky2019,Jiang2019,Choi2019,Lisi2020}. These bands are gaped out from the remaining electron energies, and, together with a reasonably large phonon frequency\,\cite{Choi2018,Cocemasov2013} that has been shown to be relevant for explaining superconductivity\,\cite{Schrodi2020}, constitute an ideal testing ground for the phenomenon discovered here. Although the predicted intensity of the replica bands is relatively weak, we propose here their detection in TBG in the foreseeable future. 
{Notably, the here-predicted replica bands due to EPI are distinct from the recently reported split-off peaks, observed in STS measurements \cite{Xie2019}, that were attributed to electron-electron interactions which are expected to be pronounced in TBG \cite{Cao2018_1}.}

\textit{Methodology.} We consider a single-branch isotropic Einstein phonon spectrum with characteristic frequency $\Omega$. For simplicity we also treat the electron-phonon scattering strength 
$g_0$ as momentum independent. Adopting the Eliashberg formalism in imaginary frequency space, with fermion frequencies $\omega_m=\pi T(2m+1)$ at temperature $T$, we decompose the inverse electron Green's function as 
\begin{align}
\hat{G}^{-1}_l(\mathbf{k},i\omega_m) =  & i\omega_mZ(i\omega_m) \hat{\rho}_0  - \phi(i\omega_m) \hat{\rho}_1  \nonumber\\
& - (\xi_l(\mathbf{k})+\chi(i\omega_m))\hat{\rho}_3  ~. \label{invgreensfunction}
\end{align}
Above, $\xi_{l}(\mathbf{k})$ are electron energies at momentum $\mathbf{k}$ in band $l$, and we work in Nambu space\,\cite{Nambu1960} with Pauli matrices $\hat{\rho}_i$. The mass enhancement $Z(i\omega_m)$, superconducting order parameter $\phi(i\omega_m)$ and chemical potential $\chi(i\omega_m)$ do not acquire any momentum dependence due to the isotropic nature of the electron-phonon coupling. By using the electron self-energy
\begin{align}
\hat{\Sigma}(i\omega_m) =& -T \sum_{\mathbf{k},m',l'} \hat{\rho}_3 \hat{G}_{l'}(\mathbf{k},i\omega_{m'}) \hat{\rho}_3  g_0^2  D(iq_{m-m'}) , \label{sigma}
\end{align}
with phonon propagator $D(iq_{m-m'})=D^0(iq_{m-m'})=-2\Omega/(\Omega^2+q_l^2)$ and boson frequencies $q_n=2\pi Tn$, we derive a closed set of equations for $Z(i\omega_m)$, $\chi(i\omega_m)$, and $\phi(i\omega_m)$ (see Appendix A). 

For the sake of clarity, we note that the here-investigated phenomenon cannot be described within the Bardeen-Cooper-Schrieffer (BCS) model, which assumes a frequency-independent order parameter, absence of mass renomalization, and an electronic bandwidth $W$ that is much larger than the phonon frequency, $W \gg \Omega$.
 Hence, we employ full-bandwidth Eliashberg theory which explicitly takes into account scattering processes involving electrons with energies and momenta that are not restricted to the vicinity of the Fermi surface, and therefore goes beyond
Migdal's theorem \cite{Schrodi2020_2}. Moreover, despite the fact that $\Omega$ is much larger than the Fermi energy and thus the systems under study lie in the extreme antiadiabatic limit, we show explicitly in the Appendix B that vertex corrections are negligible for the systems studied here.

Our self-consistent results in Matsubara space are analytically continued to the real-frequency axis, $i\omega_m\rightarrow \omega+i\delta$. This step is carried out in a formally exact and self-consistent manner, via the method first introduced by Marsiglio \textit{et al.}\,\cite{Marsiglio1988} and extended for finite electron energy bandwidths as done in Ref.\,\cite{Aperis2018}, resulting in functions $Z(\omega)$, $\chi(\omega)$, and $\phi(\omega)$. With the real-frequency dependent electron Green's function at hand we can calculate the band and momentum resolved spectral function
\begin{align}
A_l(\mathbf{k},\omega)&=-\frac{1}{\pi} \mathrm{Im}\left( \big[ \hat{G}_l(\mathbf{k},\omega+i\delta) \big]_{11} \right) \label{spectral} ,
\end{align}
which can be compared to ARPES measurements when summed over band index $l$. By further summing up the momentum degree of freedom we have the means to compare our calculations to tunneling experiments,
\begin{align}
\frac{\mathrm{d}I}{\mathrm{d}V} &\propto A(\omega)= \sum_{\mathbf{k},l} A_l(\mathbf{k},\omega) ~.\label{didv}
\end{align}
All our calculations are performed with the Uppsala Superconductivity  (\textsc{uppsc}) code\, \cite{UppSC,Aperis2015,Aperis2018,Bekaert2018,Schrodi2020_2,Schrodi2020_3}. For further details on the theory see also 
\cite{Aperis2018,Schrodi2020} and the Appendix.

\textit{Results}. We begin with a conceptually rather easy case of a nearest neighbor, one band tight-binding model on a 2D square lattice with bare electron energies $\xi(\mathbf{k})=-2t [\cos(k_x)+\cos(k_y)]-\mu$ (we drop the band index). The hopping energies and chemical potential respectively are fixed at $t=0.425\,\mathrm{meV}$ and $\mu=-1\,\mathrm{meV}$. Unless specified otherwise, we choose the electron-phonon scattering strength as $g_0=2\,\mathrm{meV}$ and a relatively large phonon frequency $\Omega=11\,\mathrm{meV}$. With an electronic bandwidth of $W=3.4\,\mathrm{meV}$ the energies $\xi(\mathbf{k})$ appear flat when compared to the phonon energy scale. We are not primarily interested in superconductivity, so, unless noted otherwise, we consider here $T>T_c$ corresponding to $\phi(\omega)=0$. The replica bands under discussion occur outside the electron bandwidth of the flat band, hence not in a frequency regime where the superconducting energy gap alters the spectrum. Therefore superconductivity does not play any role for the current analysis, as we show in the Appendix C.
 With this input we solve the full-bandwidth Eliashberg equations in Matsubara space and analytically continue the results to the real-frequency axis.

\begin{figure*}[t!]
	\centering
	\includegraphics[width=1\textwidth]{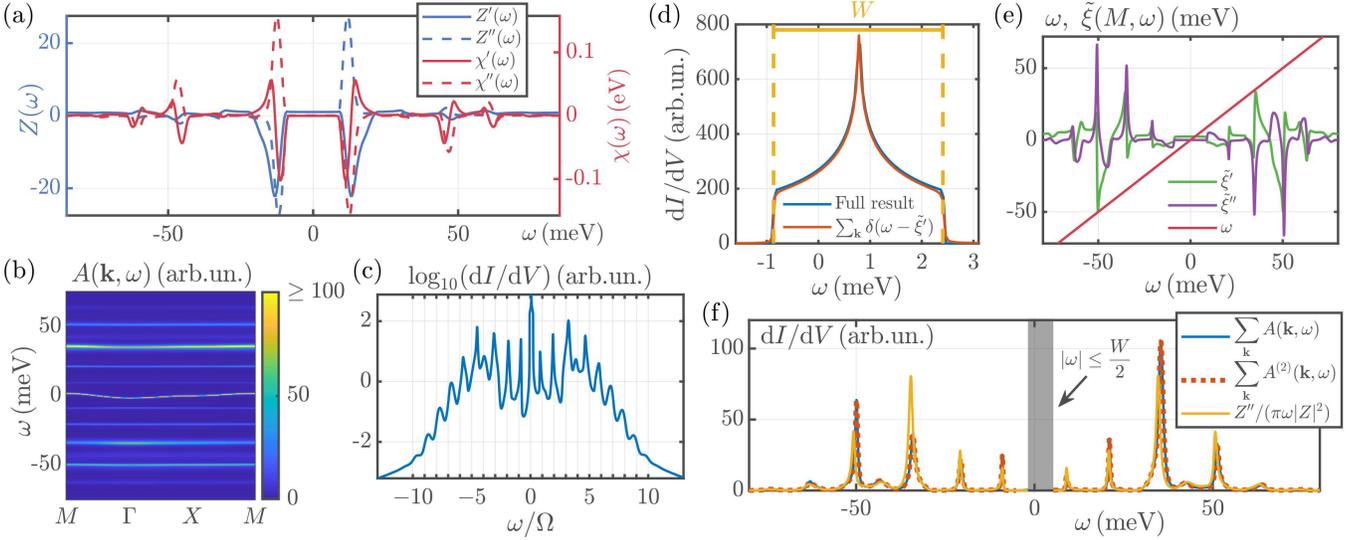}
	\caption{Real-frequency dependent results from our  self-consistent Eliashberg calculations, computed for $g_0=2\,\mathrm{meV}$, $\Omega=11\,\mathrm{meV}$ and $T>T_c$. (a) Mass enhancement and chemical potential renormalization are shown in blue and red, respectively. (b) Frequency dependent spectral function along high-symmetry lines of the 2D  tetragonal BZ, showing the replica bands. (c) Logarithmic differential conductance. (d) Differential conductance for $|\omega|\leq W/2$; the blue curve represents our full result and the red line is found from a {sum} of delta functions. (e) Renormalized electron dispersion at {$\mathbf{k}=(\pi,\pi)=M$}, the real and imaginary part are plotted as green and purples curves. In red we show $\omega$ as guide for the eye. (f) Differential conductance with focus on $|\omega|> W/2$. The three curves correspond to different ways of calculating $\mathrm{d}I/\mathrm{d}V$,  see legend and main text.}	\label{toymodel_res}
\end{figure*}

In Fig.\,\ref{toymodel_res}(a) we show the self-consistent results for the mass renormalization $Z(\omega)=Z^{\prime}(\omega)+iZ^{\prime\prime}(\omega)$ in blue and chemical potential $\chi(\omega)=\chi^{\prime}(\omega)+i\chi^{\prime\prime}(\omega)$ in red. Real and imaginary parts for both functions are drawn as solid and dashed lines, respectively. Within the electronic bandwidth, $|\omega|\leq W/2$, the real part of the mass renormalization takes on values close to unity, while $\chi^{\prime}$ is in the range of $\mu\mathrm{eV}$. This behavior is somewhat expected due to the relatively small coupling $\lambda=2g_0^2N_0/\Omega\simeq0.175$ ($N_0$: density of states at the Fermi level) and electron energy scale. In this frequency range $Z^{\prime\prime}(\omega)$ and $\chi^{\prime\prime}(\omega)$ are negligible. Turning to $|\omega|>W/2$, we observe a highly unexpected behavior of both $Z(\omega)$ and $\chi(\omega)$. Apart from unusually large magnitudes in all four functions plotted in Fig.\,\ref{toymodel_res}(a), we find large negative values for the mass renormalization at various frequencies. Although not straight-forward to physically interpret, it has been shown by Marsiglio and Carbotte that such values for $Z(\omega)$ can occur in the very strong coupling limit\,\cite{Marsiglio1991}.

Next we look into the ARPES spectrum, which we compute from Eq.\,(\ref{spectral}) with a smearing of $\delta=0.01\,\mathrm{meV}$. Our result for $A(\mathbf{k},\omega)$ is shown along high-symmetry lines of the tetragonal Brillouin zone (BZ), and as function of frequency in Fig.\,\ref{toymodel_res}(b). Here it is apparent that multiple nearly flat quasiparticle bands occur below and above the Fermi level. The frequencies corresponding to these rather coherent features seem to be separated by approximately the Einstein phonon frequency. To more reliably examine the energy positions of the observed replica bands, we show the logarithmic differential conductance in Fig.\,\ref{toymodel_res}(c), where the frequency axis is normalized to $\Omega$. The highest peak is observed at $\omega/\Omega=0$ and corresponds to the one band electron dispersion. The next three peaks to both the left and right occur very accurately at multiples of $\Omega$. All remaining signals can also be attributed to frequencies $p\,\Omega$ with $p\in\mathbb{Z}$, but slightly shifted. Empirically we therefore conclude that a cascade of quasiparticle bands 
centered at $p\,\Omega$ exists, which represent replications of the original energy band. We provide a more rigorous proof of this argument in Appendix D  \ref{appScPhonon}
, where we show that the positions of replica bands directly follow changes in $\Omega$.

It is worthwhile investigating which part of $A(\mathbf{k},\omega)$ is responsible for producing the replica bands. For this purpose we use Eq.\,(\ref{invgreensfunction}) to write the spectral function as:
\begin{eqnarray}\label{retgreen}
A(\mathbf{k},\omega)&=&-\frac{1}{\pi}\textrm{Im}\left\{\frac{1}{Z(\omega)}\frac{1}{\omega+i\delta-\tilde{\xi}(\mathbf{k},\omega)}\right\}\, ,
\end{eqnarray}
with $\tilde{\xi}(\mathbf{k},\omega)=(\xi(\mathbf{k})+\chi(\omega))/Z(\omega)$ the renormalized electron energy dispersion. To make further progress, we analyze $A(\mathbf{k},\omega)$ for two different frequency regions: In one case, $A^{(1)}$, the frequency lies within the bandwidth, $|\omega|\leq W/2$, and in the other case, $A^{(2)}$, we consider $|\omega|>W/2$. 
 
\underline{\em Case $|\omega|\leq W/2$:}
 From our numerical results we find that the imaginary part $\tilde{\xi}''(\mathbf{k},\omega)$ of the renormalized dispersion is to first order negligible in this frequency region. This is due to the fact that for $|\omega|\leq W/2$, $Z(\omega)=Z'(\omega)+iZ''(\omega)\simeq Z'(\omega)$ and $\tilde{\chi}''(\mathbf{k},\omega)\approx 0$, as well. The spectral function found from Eq.\,(\ref{retgreen}) in this case is
\begin{eqnarray}\label{a1}
A^{(1)}(\mathbf{k},|\omega|\leq W/2)&\simeq&\frac{1}{Z'(\omega)}\delta(\omega-\tilde{\xi}'(\mathbf{k},\omega)) ,
\end{eqnarray}
and corresponds to the coherent part of the quasiparticle excitation spectrum. Since $Z'(\omega)$ is nearly constant for the frequencies under consideration, and its value is close to unity, we set for simplicity $Z^{\prime}(\omega)=1$. Under this assumption Eq.\,(\ref{a1}) simplifies to 
$A^{(1)}(\mathbf{k},|\omega|\leq W/2) \simeq \delta(\omega-\tilde{\xi}'(\mathbf{k},\omega))$, so that the corresponding differential conductance can be approximated as 
\begin{align}
\frac{\mathrm{d}I^{(1)}}{\mathrm{d}V}(|\omega|\leq W/2) \simeq \sum_{\mathbf{k}} \delta(\omega-\tilde{\xi}'(\mathbf{k},\omega))~. 
\label{akwvs}
\end{align}
To show that Eq.\,(\ref{akwvs}) provides the main contributions to the tunneling spectrum for $|\omega|\leq W/2$, we plot the result as red curve in Fig.\,\ref{toymodel_res}(d). The delta-function is approximated as $\delta(x)\simeq \exp(-x^2/(2\sigma^2))/\sqrt{2\pi\sigma^2}$ with smearing $\sigma=0.01\,\mathrm{meV}$. For comparison, $\mathrm{d}I/\mathrm{d}V$, as obtained by summing the non-simplified Eq.\,(\ref{retgreen}) over momenta, is drawn in blue. The boundaries of the renormalized electron dispersion $\tilde{\xi}'(\mathbf{k},\omega)$ are indicated in yellow. We see that, despite the approximations made, the sum over delta-functions in Eq.\,(\ref{akwvs}) reproduces the full spectrum to very high accuracy for $|\omega|\leq W/2$.  The spectral features for frequencies outside the electronic bandwidth, see Fig.\,\ref{toymodel_res}(b), are due to the incoherent part of $A(\mathbf{k},\omega)$ as we will show in the following. 
 
\underline{\em Case $|\omega|> W/2$:} In this region, none of the imaginary parts in Eq.\,(\ref{retgreen}) are negligible, i.e., $Z''(\omega)\neq 0$, $\chi''(\omega)\neq 0$, therefore the spectral function has the general form,
\begin{align}
A^{(2)}(\mathbf{k},|\omega|>W/2) &= \frac{1}{\pi} \frac{\omega Z''(\omega)-\chi''(\omega)}{|\omega Z(\omega) - [\xi(\mathbf{k})+\chi(\omega)]|^2} ~,     \label{a2}
\end{align}
and refers to the non-coherent quasiparticle contributions. To prove this, we show in Fig.\,\ref{toymodel_res}(e) the real and imaginary part of $\tilde{\xi}(\mathbf{k}=M,\omega)$ in green and purple, respectively. The relation $f(\omega)=\omega$ is shown in red. We see that the only intersections of $\omega$ with $\tilde{\xi}^{\prime}$ are close to the Fermi level, i.e., for $|\omega|\leq W/2$. In this region, Eq.\,(\ref{a2}) contains sharp poles since $\omega\simeq\tilde{\xi}'(\mathbf{k},\omega)$ and $\tilde{\xi}''(\mathbf{k},\omega)\simeq 0$ as discussed in the case for $|\omega|\leq W/2$. For the two frequencies outside the electron bandwidth, i.e.\ near $-50\,\mathrm{meV}$ and $40\,\mathrm{meV}$, where $\omega\approx\tilde{\xi}'$, the imaginary part $\tilde{\xi}^{\prime\prime}$ is clearly non-zero. Hence their spectral signatures are suppressed. 

We can write the differential conductance, which results from $A^{(2)}(\mathbf{k},\omega)$, as
\begin{align}
\frac{\mathrm{d}I^{(2)}}{\mathrm{d}V}(|\omega|> W/2)=\sum_{\mathbf{k}} \frac{1}{\pi} \frac{\omega Z''(\omega)-\chi''(\omega)}{|\omega Z(\omega) - [\xi(\mathbf{k})+\chi(\omega)]|^2} . \label{didv2}
\end{align}
The outcome of Eq.\,(\ref{didv2}) is shown in Fig.\,\ref{toymodel_res}(f) as dotted red curve. The full differential conductance as obtained from Eq.\,(\ref{didv}) is plotted in solid blue. We observe that the two curves fall precisely on top of each other, which shows explicitly that the replicas originate from $A^{(2)}$ only, and no contribution from $A^{(1)}$ enters for $|\omega|> W/2$. As it turns out, we can reproduce the main features of the large-frequency spectrum by further simplifying Eq.\,(\ref{didv2}). First, since $|\omega|> W/2$, we can assume that $\xi(\mathbf{k})$ has a comparatively minor influence on both, peak positions and amplitudes. Second, we might set the chemical potential renormalization to zero, which is a rather drastic simplification since neither $\chi'(\omega)$ nor $\chi''(\omega)$ are negligibly small for $|\omega|> W/2$, compare Fig.\,\ref{toymodel_res}(a). However, assuming that these simplifications are valid we can write
\begin{align}
A^{(2)}(\mathbf{k},|\omega|> W/2) & \simeq \frac{1}{\pi\omega} \frac{Z''(\omega)}{|Z(\omega)|^2} \propto \frac{\mathrm{d}I^{(2)}}{\mathrm{d}V}(\omega) ~. \label{a2vs}
\end{align}
Note, that there is no longer a momentum dependence in Eq.\,(\ref{a2vs}) since we neglect $\xi(\mathbf{k})$. The outcome of the above expression is shown in Fig.\,\ref{toymodel_res}(f) in solid yellow. It is directly evident that the peak positions agree well with the full solution shown as blue curve. The heights do not precisely match the reference curve, which can be understood from the neglected `$-\chi''(\omega)$' in the numerator of Eq.\,(\ref{didv2}). Therefore we can conclude that the large-$\omega$ tunneling features are mainly mediated by the mass renormalization. For obtaining the correct intensities one needs to also include the chemical potential into the calculation. The bare electron dispersion $\xi(\mathbf{k})$ plays a negligible role here.

We note that our results show similarities to a study by Marsiglio and Carbotte\,\cite{Marsiglio1991}, who investigated frequency dependent results of isotropic Eliashberg theory in the strong coupling limit. They showed the existence of quasiparticle-like excitations in the spectral function, that are located at $\Delta_0+p\,\Omega$, with $\Delta_0$ the superconducting gap edge. While we similarly find a sequence of replica bands, our results differ in that we are not depending on the limit $\lambda\rightarrow\infty$, and the features detected in this work are independent of superconductivity, see Appendix.
 We further note that our itinerant Eliashberg theory results bear similarities with the well-known sequence of spectral peaks that arise in the strongly localized limit \cite{mah00}. However, as we show below,  our replica bands exist even when the electron bandwidth is the largest energy scale and most importantly, they stem from the strong mass renormalization effects that are absent in the former case.

\begin{figure}[t!]
	\centering
	\includegraphics[width=0.9\linewidth]{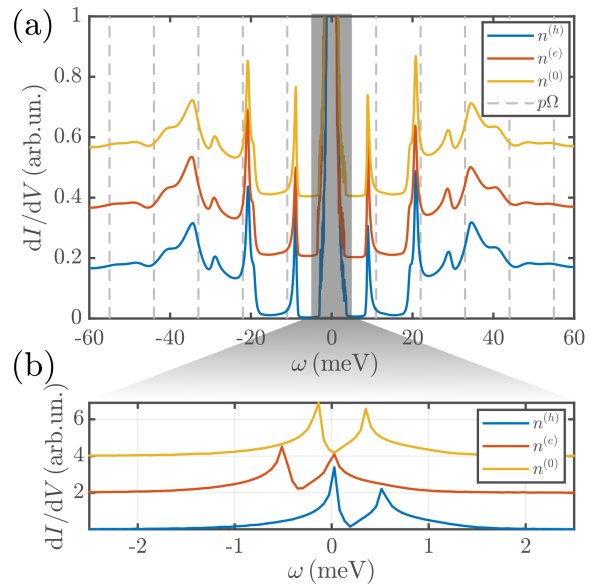}
	\caption{Tunneling spectrum computed for the four-band model of TBG. Results for different electron filling $n$ have been shifted vertically with respect to each other. (a) High-frequency contributions for fillings $n^{(h)}$ (blue), $n^{(e)}$ (red) and $n^{(0)}$ (yellow). (b) Low-frequency spectrum for the same fillings and colors as in (a).}	\label{tbg_tun}
\end{figure}

\textit{Twisted bilayer graphene}. 
Let us now turn to TBG at a twist angle of $\sim1.1^{\circ}$, where we fix the phonon frequency at $\Omega=11\,\mathrm{meV}$\,\cite{Choi2018,Cocemasov2013} and use a faithful {ten-band} tight-binding model for the electron energies\,\cite{Po2018_2}. {This model has two flat bands near the Fermi level with a narrow bandwidth $W\approx 7$ meV which are energetically separated from the rest of the bands by energy gaps over 20 meV.} Further, we consider the normal state, $T=1.6\,\mathrm{K}>T_c$\,\cite{Cao2018_2,Yankowitz2019} and choose an electron-phonon scattering strength $g_0=1.6\,\mathrm{meV}$\,\cite{Schrodi2020}. When performing our analysis for the two flat bands only, the resulting spectra are very similar to our model calculations above and we present the outcomes in  the Appendix E.
 Next, we take into account a total of four energy bands with bandwidth $W\simeq127\,\mathrm{meV}>\Omega$. This includes the two flat bands close to the Fermi level, enclosing an energy window of around $7\,\mathrm{meV}<\Omega$, and an additional occupied and unoccupied band below and above. In Fig.\,\ref{tbg_tun} we show $\mathrm{d}I/\mathrm{d}V$ for three different electron fillings $n$. Here, $n^{(0)}$ corresponds to half filling and for $n^{(e)}$ ($n^{(h)}$) the Fermi level lies exactly at the van Hove singularity of the unoccupied (occupied) band of the bare system. As apparent in panel (a), the replicas are superimposed with the additional non-flat energy bands, which are located at approximately $|\omega|\gtrsim20\,\mathrm{meV}$. However, the signals at $p=\pm1,\pm2,\pm3$ are still clearly resolvable. The outcomes for the low-$\omega$ regime are drawn in Fig.\,\ref{tbg_tun}(b). A closer inspection of both panels reveals that the filling does not noticeably influence the intensity or location of the replicas, which is expected due to the comparatively small energy scale {on} which $\xi_l(\mathbf{k})$ is shifted.

{Based on these findings, we are confident to predict the observation of signals at $\omega=p\,\Omega$ in experiment.} Since the replica bands do not represent coherent excitations of the system, their intensity is significantly lower than for actual poles of the Green's function (by about a factor $\mathcal{O}(50)$). Consequently, the proposed tunneling features are not expected to be prominent in the experimental spectra however they should be detectable, given the available resolution \cite{Machida2019}. The observation of the here-predicted quasiparticle replica bands for TBG would provide strong support for {the importance of the electron-phonon interaction} for its low-temperature behavior. 

Our predicted replicas are distinctly different from those observed in FeSe/STO \cite{Lee2014} that are due to the forward-scattering (small-q) EPI, whereas the here considered flat-band systems have a momentum independent EPI. Since in the latter case the phenomenon is driven by the electron flat bands themselves, the complete main bands are replicated, above and below the Fermi energy, and not only a part of an electron band  below the Fermi energy as in the case of FeSe/STO \cite{Rademaker2016,Aperis2018}. 
Moreover, the flat-band replicas have spectral intensity which is maximal at the second replica and they appear in multiplets, i.e.\ in a cascade of several replica bands. In contrast, for small-q EPI mediated replicas almost all the spectral intensity is found in the first replica band \cite{Rademaker2016,Schrodi2018}. Although the flat-band replicas are weak in intensity, they should be more pronounced than those of FeSe/STO. 

\textit{Conclusions.}
To summarize, we predict a sequence of quasiparticle replicas in 2D systems that exhibit both, one or more flat bands around the Fermi level, and a comparatively large phonon energy scale. The conditions necessary to observe these features in ARPES or STS/STM measurements are a high experimental accuracy, and sufficient energy gaps between the flat band(s) and the remaining (un)occupied levels. The prime candidate for detecting such signals is TBG, where the flat bands close to the Fermi energy are isolated to a good approximation. Our calculations explicitly show that the replicas, occurring at multiples of the phonon frequency, are well distinguishable from the spectral signals of neighboring energy bands. Although TBG has been studied extensively, no such sequence of replicas has yet been discussed or observed experimentally, up to our knowledge. 
Most experiments focus on a frequency range comparable to the bandwidth of the two flat bands, while for the observation of the here-discovered phenomenon an investigation of frequencies at least up to $\sim50\,\mathrm{meV}$ is required. Lastly, split-off peaks were recently observed in STS measurements on TBG that were attributed to strong electron-electron correlations \cite{Xie2019}. Our predicted replicas have a distinct origin, but appear as similar STS peaks whose position and magnitude depends on the mediating EPI. Hence, we predict that the quasiparticle spectrum of TBG harbors more surprises in the form of EPI-mediated replica bands. Given the generality of our analysis, our predictions should be relevant to all related flat band systems \cite{Heikkilae2011,Leykam2018,Balents2020}.

\begin{acknowledgments}
This work has been supported by the Swedish Research Council (VR), the R{\"o}ntgen-{\AA}ngstr{\"o}m Cluster and the Knut and Alice Wallenberg Foundation (grant No.\ 2015.0060).The calculations were enabled by resources provided by the Swedish National Infrastructure for Computing (SNIC) at NSC Link\"oping, partially funded by the Swedish Research Council through grant agreement No.\ 2018-05973.	
\end{acknowledgments}

\appendix

\section{Eliashberg theory}\label{appScEq}

We model our physical system by the Hamiltonian
\begin{align}
H =& \sum_{\mathbf{k},l} \xi_{l}(\mathbf{k}) \Psi^{\dagger}_{l}(\mathbf{k}) \hat{\rho}_3 \Psi_{l}(\mathbf{k}) + \hbar\Omega \sum_{\mathbf{q}} \Big(b^{\dagger}(\mathbf{q})b(\mathbf{q}) + \frac{1}{2}\Big) \nonumber\\
& + g_0\sum_{\mathbf{k},\mathbf{k}',l,l'} u(\mathbf{k}-\mathbf{k}') \Psi^{\dagger}_{l'}(\mathbf{k}') \hat{\rho}_3 \Psi_l(\mathbf{k})
\end{align}
with phonon displacement $u(\mathbf{q})=b(\mathbf{q})+b^{\dagger}(-\mathbf{q})$ and Nambu spinor $\Psi_l(\mathbf{k})=\big(c^{\dagger}_{l,\uparrow}(\mathbf{k}), c_{l,\downarrow}(-\mathbf{k})\big)$. Here we use $b^{\dagger}(\mathbf{q})$ and $c^{\dagger}_{l,\sigma}(\mathbf{k})$ as phonon and electron creation operators, with $\sigma\in\{\uparrow,\downarrow\}$ as spin label. The phonon frequency $\Omega$ and electron-phonon scattering elements $g_0$ are both assumed to be isotropic. The electron Green's function $\hat{G}_l(\mathbf{k},i\omega_m)$, as defined in Eq.\,(\ref{invgreensfunction}) of the main text, obeys the Dyson equation
\begin{align}
\hat{G}_l(\mathbf{k},i\omega_m) = \hat{G}^0_l(\mathbf{k},i\omega_m) \hat{\Sigma}(i\omega_m) \hat{G}_l(\mathbf{k},i\omega_m) ~, \label{dyson}
\end{align}
with $\hat{\Sigma}(i\omega_m)$ given by Eq.\,(\ref{sigma}) of the main text. The non-interacting Green's function in Eq.\,(\ref{dyson}) is $\big[\hat{G}^0_l(\mathbf{k},i\omega_m)\big]^{-1} = i\omega\hat{\rho}_0 - \xi_l(\mathbf{k})\hat{\rho}_3$, so that we obtain
\begin{align}
~~&\hat{G}_l(\mathbf{k},i\omega_m) = \big[i\omega_mZ(i\omega_m) \hat{\rho}_0  + \phi(i\omega_m) \hat{\rho}_1  \nonumber\\
&~~~~~~~~~~+ (\xi_l(\mathbf{k})+\chi(i\omega_m)))\hat{\rho}_3 \big] \Theta^{-1}_l(\mathbf{k},i\omega_m) ~, \label{greensfunction}\\
&\Theta_l(\mathbf{k},i\omega_m) = (i\omega_mZ(i\omega_m))^2 - \phi^2(i\omega_m)   \nonumber\\
&~~~~~~~~~~- (\xi_l(\mathbf{k})+\chi(i\omega_m))^2 ~.
\end{align}

As we describe in the main text, the phonon propagator is approximated by $D(\mathbf{q},iq_n)=D^0(\mathbf{q},iq_n)$, so that we get the electron-phonon interaction kernel
\begin{align}
V^{\mathrm{e-ph}}(iq_{n}) = g_0^2 \frac{2\Omega}{\Omega^2 + q_n^2} ~.
\end{align}
The resulting Eliashberg equations for the mass renormalization $Z(i\omega_m)$, chemical potential $\chi(i\omega_m)$ and superconducting order parameter $\phi(i\omega_m)$ in Matsubara space read
\begin{align}
Z(i\omega_m) &= 1-\frac{T}{\omega_m} \sum_{\mathbf{k},m',l} V^{\mathrm{e-ph}}(iq_{m-m'})\frac{\omega_{m'}Z(i\omega_{m'})}{\Theta_l(\mathbf{k},i\omega_{m'})}, \label{z} \\
\chi(i\omega_m) &= T \sum_{\mathbf{k},m',l} V^{\mathrm{e-ph}}(iq_{m-m'}) \frac{\xi_l(\mathbf{k})+\chi(i\omega_{m'})}{\Theta_l(\mathbf{k},i\omega_{m'})},\\
\phi(i\omega_m) &= -T \sum_{\mathbf{k},m',l} V^{\mathrm{e-ph}}(iq_{m-m'})\frac{\phi(i\omega_{m'})}{\Theta_l(\mathbf{k},i\omega_{m'})} ~. \label{phi}
\end{align}
The electron filling of the system is given by
\begin{align}
n = 1 - \frac{2T}{L}\sum_{\mathbf{k},m}\sum_l \frac{\xi_l(\mathbf{k})+\chi(i\omega_m)}{\Theta_l(\mathbf{k},i\omega_m)} \label{n1}
\end{align}
where $L$ denotes the number of electronic bands.

Once we have solved the Eliashberg equations in Matsubara space, we can calculate ARPES and STM spectra to make direct contact with experiment\,\cite{Aperis2018}. For this purpose we analytically continue the solutions to Eqs.\,(\ref{z}-\ref{phi}) self-consistently via
\begin{align}
&Z(\omega) = 1 - \frac{T}{\omega}\sum_{\mathbf{k},m,l} V^{\mathrm{e-ph}}(\omega-\omega_m) \frac{i\omega_mZ(i\omega_m)}{\Theta_l(\mathbf{k},i\omega_m)} \nonumber  \\
&-\frac{1}{2\omega}\int_{-\infty}^{\infty} \mathrm{d}z\sum_{\mathbf{k},l} \frac{\alpha^2F_l(z)}{N_l(0)} \frac{Z(\omega-z)(\omega-z)}{\Theta_l(\mathbf{k},\omega-z)} \zeta(\omega,z) , \label{zomega}\\
&\chi(\omega) = T\sum_{\mathbf{k},m,l} V^{\mathrm{e-ph}}(\omega-\omega_m) \frac{\xi_l(\mathbf{k})+\chi(i\omega_m)}{\Theta_l(\mathbf{k},i\omega_m)}\nonumber \\
&+\int_{-\infty}^{\infty} \mathrm{d}z\sum_{\mathbf{k},l} \frac{\alpha^2F_l(z)}{2N_l(0)} \frac{\xi_l(\mathbf{k})+\phi(\omega-z)}{\Theta_l(\mathbf{k},\omega-z)} \zeta(\omega,z) ~,~~\\
&\phi(\omega) = -T\sum_{\mathbf{k},m,l} V^{\mathrm{e-ph}}(\omega-\omega_m) \frac{\phi(i\omega_m)}{\Theta_l(\mathbf{k},i\omega_m)} \nonumber \\
& +\int_{-\infty}^{\infty} \mathrm{d}z\sum_{\mathbf{k},l} \frac{\alpha^2F_l(z)}{2N_l(0)} \frac{\phi(\omega-z)}{\Theta_l(\mathbf{k},\omega-z)} \zeta(\omega,z) , \label{phiomega}
\end{align}
with $\zeta(\omega,z)=\left(\tanh\frac{\omega-z}{2T}+\coth\frac{z}{2T}\right)$ introduced for brevity\, (see \cite{Marsiglio1988,Aperis2018}), and $N_l(0)$ the band-resolved density of states at the Fermi level. The solutions to Eqs.\,(\ref{zomega}-\ref{phiomega}) are then used to compute the real-frequency matrix Green's function of the system, which in turn can be employed to find the momentum, band and frequency resolved spectral function as well as the STS/STM spectrum via Eqs.\,(3),\,(4) of the main text.

For solving Eqs.\,(\ref{z}-\ref{phi}) and Eqs.\,(\ref{zomega}-\ref{phiomega}) we do not make use of any further simplifications and keep the full complexity of the problem. Momentum and frequency grids have been checked for convergence. Our efficient implementation \cite{UppSC} makes use of fast Fourier transform (FFT) convolution schemes, and we exploit the known functional form of the interaction kernel to reach faster convergence in the number of Matsubara frequencies\,\cite{Schrodi2019}.

\section{Vertex corrections}\label{beyondMigdal}

The flat band systems that we are concerned with here, lie in the extreme antiadiabatic regime where the phonon energy is much larger than the Fermi energy, i.e. $\Omega>>E_F$. As such, it is reasonable to expect violations of Migdal's theorem. Generally, corrections beyond Migdal's theorem involve two kind of contributions: i) corrections due to the finite bandwidth that are not accounted for due to the Fermi surface based derivation of the theorem and ii) corrections to the electron-phonon vertex beyond lowest order \cite{Danylenko2001,Schrodi2020_2}. We stress that the former are fully taken into account in our full-bandwidth Eliashberg theory, hence our approach goes beyond standard Migdal-Eliashberg theory \cite{Schrodi2020_2}. Here, we will show that the latter, i.e. vertex corrections, are negligible for the systems that we study here.

For simplicity, we consider a model system with one band that is perfectly flat, i.e. $\xi_{\bf k}=\epsilon$. This approximation is reasonable since for our model system discussed in the main text $\Omega>>W$. It also renders the problem tractable analytically to a large extent. The corresponding vertex function in the one-loop approximation, i.e. in terms of the the normal state  non-interacting Green's function, has the form (see e.g.  \cite{Schrodi2020_2}):
\begin{eqnarray}\no
\Gamma(i\omega_{m},i\omega_{m'})&=&T\sum_{{\bf k}'',m''}g_0^2 D(\omega_{m'}-\omega_{m''})G_0(i\omega_{m''})\\\no
&\times&G_0(i\omega_{m''}-i\omega_{m'}+i\omega_{m})
\end{eqnarray}  
with $G^{-1}_0(\omega_m)=i\omega_m Z(i\omega_m)-\epsilon$. Taking the $m=m'$ case, we have 
\begin{eqnarray}\no
\Gamma(i\omega_{m})&=&T\sum_{m'}\frac{2g_0^2}{\left[\Omega^2+(\omega_m-\omega_{m'})^2\right]}\\\label{flatvertex2}
&\times& \left(i\omega_{m'}Z(i\omega_{m'})\Omega^{-1}-\beta\right)^{-2}
\end{eqnarray}  
where in the last step we have introduced the ratio $\beta=\epsilon/\Omega$. To make further progress, we take a square-well model ansatz for the mass renormalization function: $Z(i\omega_n)=1+\lambda$ for $|\omega_n|<\omega_c$ and $Z(i\omega_n)=1$ elsewhere, with $\omega_c$ a frequency cutoff. 
The Matsubara sum of Eq.\,(\ref{flatvertex2}) can now be performed analytically, but the resulting expression is too lengthy and we will not report it here. We have checked numerically that the vertex correction in Eq.\,(\ref{flatvertex2}) is a smooth function of $\omega_m$ and temperature and is maximized for $T\rightarrow 0, m=1$. Therefore, for our purposes we focus here on the zero temperature, static vertex limit,
\begin{eqnarray}\no
\Gamma&=&\int_{-\omega_c}^{\omega_c}\frac{d\omega_{m'}}{2\pi}\frac{2g_0^2}{\left(\frac{i\omega_{m'}(1+\lambda)}{\Omega}-\beta\right)^2\Bigl(\Omega^2+\omega^2_{m'}\Bigl)}\\\no
&+&\int_{-\infty}^{-\omega_c}\frac{d\omega_{m'}}{2\pi}\frac{2g_0^2}{\left(\frac{i\omega_{m'}}{\Omega}-\beta\right)^2\Bigl(\Omega^2+\omega^2_{m'}\Bigl)}\\\label{staticvertex}
&+&\int_{\omega_c}^{\infty}\frac{d\omega_{m'}}{2\pi}\frac{2g_0^2}{\left(\frac{i\omega_{m'}}{\Omega}-\beta\right)^2\Bigl(\Omega^2+\omega^2_{m'}\Bigl)}\ .
\end{eqnarray}
The integrals in the above can be performed analytically.

\begin{figure}[h!]
\includegraphics[width=0.9\linewidth]{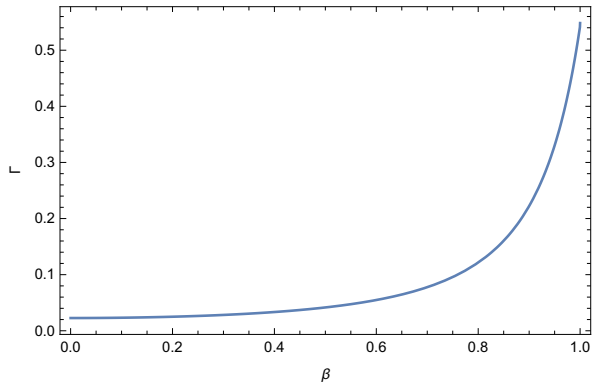}
	\caption{Computed Zero-temperature static vertex correction, $\Gamma$, as a function of the ratio $\beta=\epsilon/\Omega$.}	\label{vertexplot}
\end{figure}
Using the parameters of our model system from the main text, $\lambda=0.175$, $\Omega=11$\,meV, $g_0=2$\,meV and $\omega_c=\Omega$, we plot the above equation as a function of the ratio $\beta$ in Fig.\,\ref{vertexplot}. This figure shows clearly that for the parameter range that we are concerned with in this work ($\beta<0.3$), vertex corrections are negligible since $\Gamma<<1$.
Note that the results hardly change if we increase the cutoff, e.g. set $\omega_c=10\Omega$.

To be more specific, $\Gamma(\beta=1/11)\approx0.023$ and $\Gamma(\beta=0.23)\approx0.028$.
Near $\beta\approx 0$, a Taylor expansion of Eq.\,(\ref{staticvertex}) within our chosen parameter set yields,
\begin{eqnarray}\no
\Gamma\approx 0.023 + 0.056 \beta^2
\label{vertexTaylor}\ .
\end{eqnarray}  

\section{Results in the superconducting state}\label{appScBelowtc}

\begin{figure}[h!]
	\centering
	\includegraphics[width=0.9\linewidth]{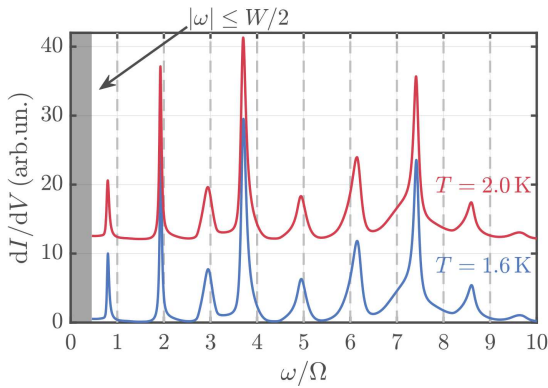}
	\caption{Tunneling spectrum computed for our model {system with bandwidth $W=3.4\,\mathrm{K}$}, choosing $g_0=3\,\mathrm{meV}$, $\Omega=11\,\mathrm{meV}$ and $T=1.6\,\mathrm{K}$ ($T=2.0\,\mathrm{meV}$) for the blue (red) curve.}	\label{scstate}
\end{figure}

As briefly mentioned in the main text, superconductivity does not play any significant role in the observation of replicated flat band(s). To explicitly prove this point we perform additional calculations for our model dispersion, using again  the phonon frequency $\Omega=11\,\mathrm{meV}$. The electron-phonon scattering strength in this section is set to $g_0=3\,\mathrm{meV}$, such that we get a maximum superconducting gap of $\Delta=\phi(0)/Z(0)\simeq215\,\mu\mathrm{eV}$ at $T=1.6\,\mathrm{K}$. The corresponding tunneling spectrum for $|\omega|> W/2$ is shown in Fig.\,\ref{scstate} as blue curve. Once we increase the temperature to $2\,\mathrm{K}$ the gap closes and we obtain results for $\mathrm{d}I/\mathrm{d}V$ as drawn in red. It is easily observed that the two spectra in Fig.\,\ref{scstate} do neither differ in the locations of replica bands, nor are the corresponding intensities visibly deviating. This confirms that superconductivity has no observable effect on the tunneling features under investigation, since the gap opening occurs only inside the electron bandwidth of the flat band.

\section{Influence of phonon frequency}\label{appScPhonon}

\begin{figure}[t!]
	\centering
	\includegraphics[width=0.9\linewidth]{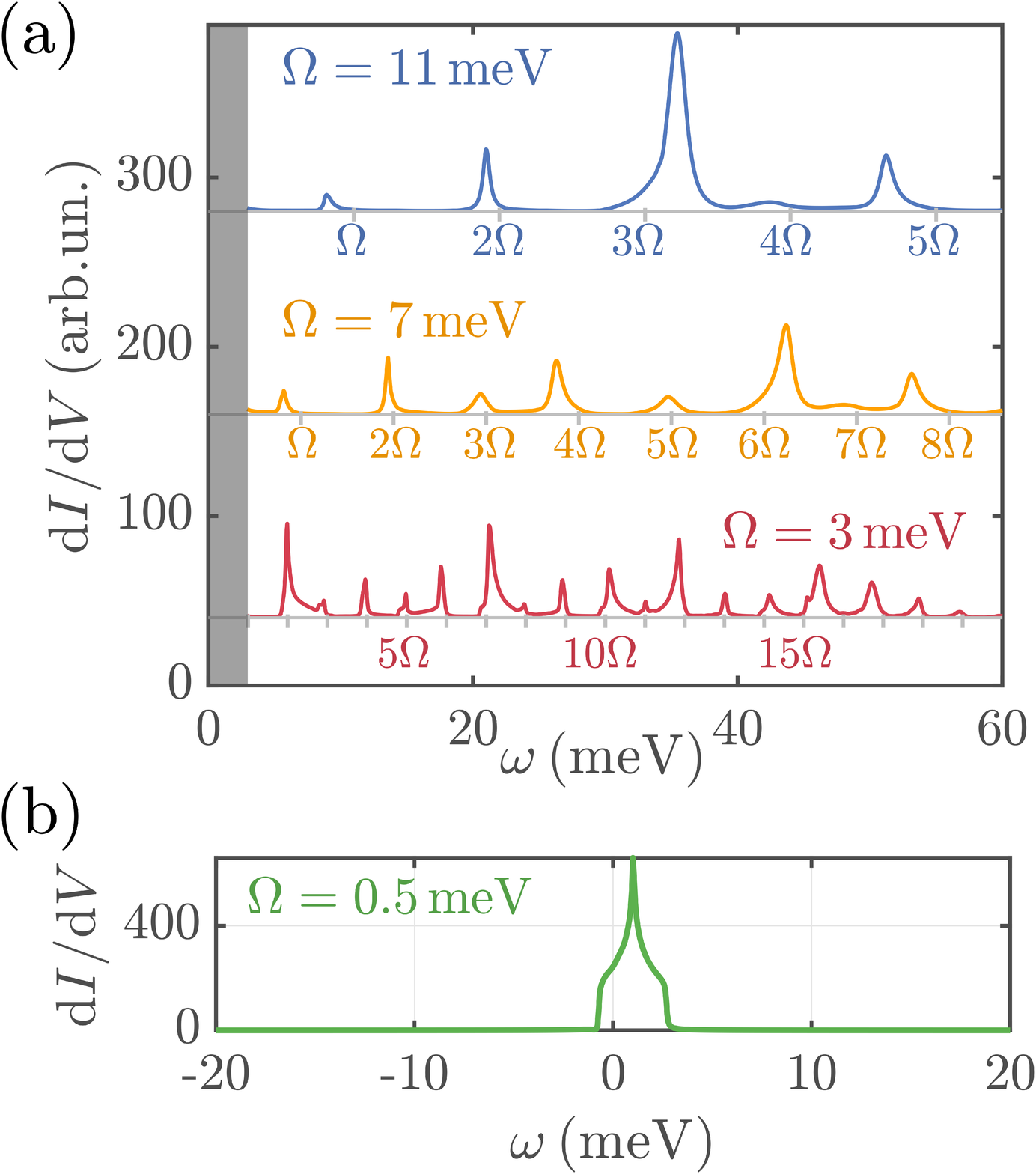}
	\caption{Tunneling spectrum as function of $\omega$, focusing on $|\omega|>W/2$ ($= 1.7$ meV). (a) Red, yellow and blue curves correspond respectively to $\Omega=3\,\mathrm{meV}$, $\Omega=7\,\mathrm{meV}$ and $\Omega=11\,\mathrm{meV}$. (b) $\mathrm{d}I/\mathrm{d}V$ as computed for $\Omega=0.5\,\mathrm{meV}$.}	\label{phononfreq}
\end{figure}

In this section we want to examine the effect of phonon frequency on our results for the model system. In the main text we chose $\Omega$ significantly larger than the electronic bandwidth {$W=3.4\,\mathrm{meV}$}, such that $\xi(\mathbf{k})$ appears as flat band in comparison. Now we additionally consider the cases where $\Omega$ is 7, 3, or 0.5\,meV, keeping $T<T_c$ and choosing the scattering strength $g_0$, such that $\lambda=2N_0g_0^2/\Omega=\mathrm{constant}$. Our results for the differential conductance are shown in Fig.\,\ref{phononfreq}(a), where the curves have been shifted vertically with respect to each other. We focus here on the large-frequency part of the tunneling spectrum and show the results only for $\omega>0$. The outcomes for $\Omega=11\,\mathrm{meV}$ (blue), $\Omega=7\,\mathrm{meV}$ (yellow) and $\Omega=3\,\mathrm{meV}$ (red) rigorously prove that the replica bands occur at multiples of the respective phonon frequencies. In Fig.\,\ref{phononfreq}(b) we draw the complete tunneling spectrum for phonon frequency $\Omega=0.5\,\mathrm{meV}$ and observe that the replica bands are absent. This behavior is to be expected, since for such small $\Omega$ the single electron band does no longer appear flat, i.e. the energy bandwidth is significantly larger than the phonon frequency.

\section{Two-band case of TBG}\label{appScTwobandstbg}

The replica bands in TBG are most easily observed when only the two flat bands close to the Fermi level are considered. In this section we therefore look into results obtained by this setup, additionally setting $T>T_c$, $\Omega=11\,\mathrm{meV}$ and $g_0=1.6\,\mathrm{meV}$. The electron bandwidth $W\simeq7\,\mathrm{meV}$ is then smaller than $\Omega$, hence we expect to observe  effects comparable to our model system in the main text and Appendix \ref{appScPhonon}. As before, we solve the Eliashberg equations in Matsubara space and analytically continue the self-consistent results to real frequencies. The ARPES and tunneling spectra are then obtained as function of $\omega$.

\begin{figure}[h!]
	\centering
	\includegraphics[width=1\linewidth]{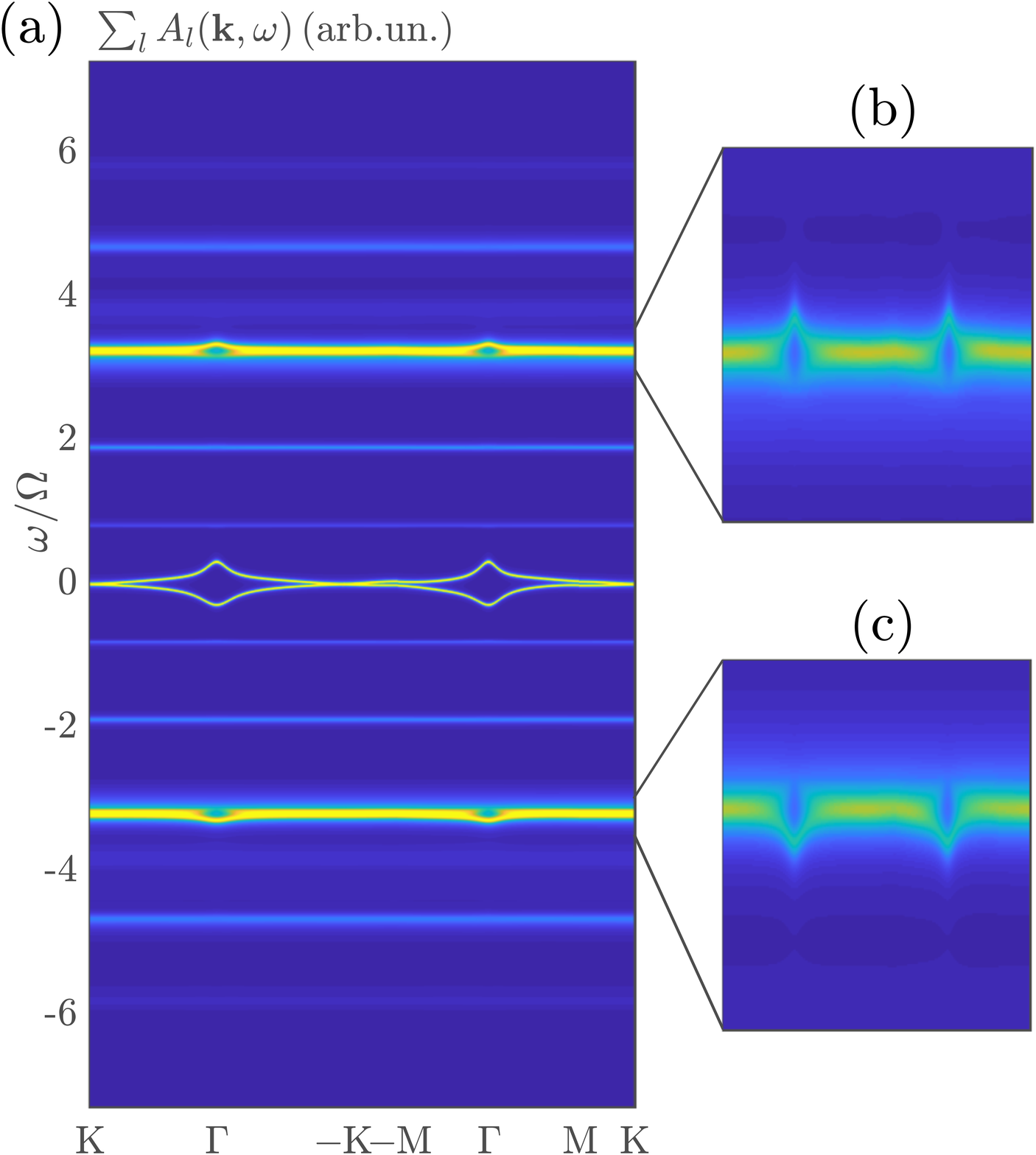}
	\caption{Momentum and frequency dependent ARPES spectrum for $n=n^{(h)}$. (a) Global view. (b) Zoom into frequency region of the most prominent replica band above the Fermi level. (c) Zoom into frequency region of the most prominent replica band below the Fermi level.}	\label{tbg2arp}
\end{figure}

\begin{figure}
\includegraphics[width=0.9\linewidth]{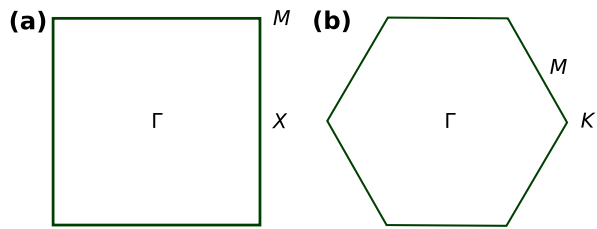}
	\caption{Brillouin Zones for the tight-binding models that we consider in this work, with high-symmetry points indicated. (a) The BZ of our model system (see Fig.\,\ref{toymodel_res} in the main text) and  (b) mini BZ of twisted bilayer graphene (see Fig.\,\ref{tbg2arp}).}	\label{BZs}
\end{figure}

In Fig.\,\ref{tbg2arp}(a) we show the spectral function, summed over energy bands, at a filling $n=n^{(e)}$ along momenta in the mini BZ of TBG (see Fig.\,\ref{BZs}) and frequencies. The bare two-band dispersion corresponds to enhanced signals close to the Fermi level. All remaining features in this graph represent replications of the original energies. As is easily seen, and in agreement to our model calculations in the main text, these signals are almost constant in momentum space and occur approximately at integer multiples of the phonon frequency. In panels (b) and (c) of Fig.\,\ref{tbg2arp} we zoom into the frequency regions around $\omega/\Omega\sim3$ and $\omega/\Omega\sim-3$, respectively. From these parts of the spectrum it is  evident that the observed features are indeed direct replications of the original $\xi_{l}(\mathbf{k})$.

\begin{figure}[h!]
	\centering
	\vspace*{0.1cm}
	\includegraphics[width=0.9\linewidth]{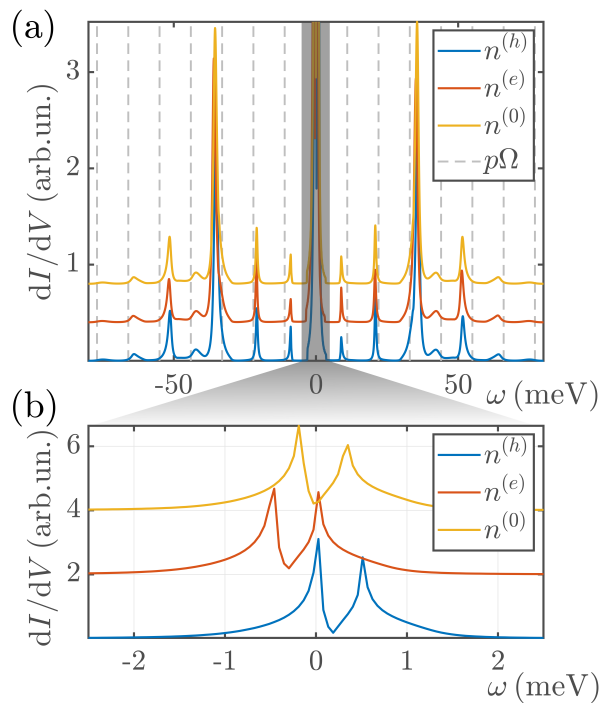}
	\caption{Tunneling spectrum computed from the two-band model of TBG, setting $T=1.6\,{\rm K}<T_c$, $g_0=1.5\,\mathrm{meV}$, $\Omega=11\,\mathrm{meV}$. Results for different $n$ have been shifted vertically with respect to each other. (a) High-frequency contributions for fillings $n^{(h)}$ (blue), $n^{(e)}$ (red) and $n^{(0)}$ (yellow). (b) Low-frequency spectrum for the same fillings and colors as in (a).}	\label{tbg2bands}
\end{figure}

We show in Fig.\,\ref{tbg2bands} the corresponding differential conductance as it can be measured by STM experiments. For better visibility we plot in Fig.\,\ref{tbg2bands}(a) the high-frequency spectra only, for fillings $n^{(h)}$ (blue), $n^{(e)}$ (red) and $n^{(0)}$ (yellow) as given in the main text. Multiples of $\Omega$ are indicated by dashed gray vertical lines. The region of small $\omega$ is shown in a magnified way in Fig.\,\ref{tbg2bands}(b) for similar fillings and color code.  We observe from this graph that the large-energy part of the spectrum contains a sequence of replicas similar to our model system. Further, the range of electron doping that is relevant for superconductivity\,\cite{Schrodi2020} does not affect the replica positions in a noticeable way, since the phonon energy scale is dominant. It should be noted that not all signals are equally well pronounced, so only some replicas could be  be observable due to experimental resolution.

\FloatBarrier

%
\end{document}